# Combined effect of SiC and carbon on sintering kinetics, microstructure and mechanical properties of fine-grained binderless tungsten carbide

E.A. Lantsev [a], P.V. Andreev [a], A.V. Nokhrin [a, (*)], Yu.V. Blagoveshchenskiy [b], N.V. Isaeva [a,b], M.S. Boldin [a], A.A. Murashov [a], G.V. Shcherbak [a], K.E. Smetanina [a], V.N. Chuvil'deev [a], N.Yu. Tabachkova [c,d]

[a] Lobachevsky State University of Nizhny Novgorod, 603022 Nizhny Novgorod, Russia

[b] A.A. Baykov Institute of Metallurgy and Materials Science of the Russian Academy of Sciences, 119334 Moscow, Russia

[c] National University of Science and Technology "MISIS", 119991 Moscow, Russia

[d] A.M. Prokhorov General Physics Institute of the Russian Academy of Sciences, 119991 Moscow, Russia

e-mail: andreev@phys.unn.ru

**Abstract:** The study investigates the density, phase composition, microstructure and mechanical properties (microhardness, fracture toughness) of binderless WC + SiC and WC + SiC + C ceramics obtained by Spark Plasma Sintering (SPS). Nanopowders of α-WC produced by DC arc plasma chemical synthesis were used as raw materials. Powder compositions for sintering contained graphite (0.3, 0.5% wt.) and β-SiC (1, 3, 5% wt.) with 0.3% wt. graphite. It was shown that WC + 1%SiC + 0.3%C ceramics have a homogeneous fine-grained microstructure, high relative density, increased microhardness and Palmquist fracture toughness (Indentation Fracture Resistance). The kinetics of the initial sintering stage of WC + C and WC + C + SiC powders was also analyzed using high-temperature dilatometry at the conventional pressureless sintering (CPS) conditions. The CPS and SPS activation energies of WC + SiC powders at the intensive shrinkage stage were determined using the Young-Cutler model. The CPS activation energies of WC, WC + C and WC + C + SiC powders are close to the activation energy of diffusion of the carbon $^{14}C$ along the α-WC grain boundaries. The SPS activation energies of WC + C and WC+ C + SiC powders turn out to be lower than the activation energy of the $^{14}$C of α-WC grain boundary.

---

(*) Corresponding author (nokhrin@nifti.unn.ru)

**Keywords:** tungsten carbide, silicon carbide, spark plasma sintering, nanopowder.

**1. Introduction**

WC–Co hard alloys with increased fracture toughness and sufficient strength are widely used in machine tool engineering for the production of metal cutting tools (MCT) [1–3]. Liquid-phase sintering is generally used to obtain WC–Co hard alloys [1, 4, 5]. This technology stands out since it ensures a uniform distribution of the low melting point metal phase in the material. New hard alloys of WC–Ni [1, 4] and WC–Fe [4] type as well as other hard alloys are produced in the same way. Fine-grained hard alloys with 6–10% Co are characterized by high hardness ($HV$ = 18–22 GPa) and Palmquist fracture toughness coefficient ($K_{IC}$ = 10–14 MPa·m$^{1/2}$).

At present, fine-grained ceramics based on binderless tungsten carbide (WC) are of great interest. The key feature of WC-based ceramics is the combination of high hardness ($HV$ > 30 GPa) and increased fracture toughness ($K_{IC}$ > 4–5 MPa·m$^{1/2}$) [6, 7]. Besides, tungsten carbide has a high melting point, high elastic modulus and high corrosion resistance. The combination of properties enables binderless tungsten carbide to be used to manufacture MCT designed for finishing engineering structural materials. However, binderless WC-based ceramics produced by conventional pressureless sintering (CPS) have low density and poor mechanical properties.

Oxides, carbides and nitrides are added to WC-based ceramics to improve the mechanical properties [5,8,9]. Recent studies suggest far-reaching prospects for dispersion-hardened WC–SiC ceramics with increased strength and hardness [10–12]. The addition of dispersed SiC particles into tungsten carbide is expected to increase its strength and fracture toughness and allow to control the friction coefficient, which is an important feature for MCT.

The second problem preventing the production of ceramics from tungsten carbide with high mechanical properties is the low stability of the α-WC phase during heating and when the carbon concentration deviates from the equilibrium value. Tungsten monocarbide (α-WC) has a very narrow stable temperature range and carbon concentration in the W–C diagram [2]. Even a slight carbon concentration deviation from the equilibrium concentration ($C_0$ = 6.13% wt.) leads to the formation of $W_2C$ particles, which reduces the mechanical properties of ceramics.

The carbon concentration in WC decreases during sintering. This is due to the fact that oxygen atoms interact with carbon atoms in the WC particles surface [2, 6]. This leads to the formation of $W_2C$ particles. Excessive carbon concentration leads to the formation of graphite particles and abnormal grain growth that also reduce the mechanical properties of WC-based ceramics [6]. Therefore, determining the optimal carbon concentrations in WC-based ceramics is a relevant task with practical implications.

DC arc plasma chemical synthesis of tungsten monocarbide nanopowders [13] and Spark Plasma Sintering (SPS) [2, 14–23] have been used to produce ceramics under study. It is assumed that the combination of these technologies will further improve the mechanical properties of WC-based ceramics due to the formation of a high-density ultrafine grained (UFG) microstructure. One of the most important SPS advantages is the possibility of sintering ceramics at high heating rates (up to 1000 °C/min) [24]. This allows limiting the grain growth and obtaining materials with a small grain size (up to 100 nm) [15, 17, 23]. Recent studies have shown promising prospects of applying plasma chemical synthesis of WC nanopowder and SPS jointly to obtain UFG ceramics based on WC with advanced mechanical properties [6, 14, 25].

The aim of this study was to produce and investigate the features of the WC–SiC ceramics sintering with varying carbon content, as well as to study the effect of sintering modes on the microstructure parameters, phase composition and mechanical properties.

## 2. Materials and Methods

### 2.1. Initial powders and raw materials

The objects of the study were binderless ceramics prepared from WC, WC + C and WC + C + SiC powder compositions. Nanopowder of tungsten carbide (WC) obtained by DC arc plasma chemical synthesis followed by hydrogen annealing at 1050°C (3 h) was used as raw material [13].

Preparation of the powder compositions WC + C and WC + C + SiC for ceramics of Group II and III was carried out in several stages. Carbon in the form of colloidal graphite (0.3, 0.5% wt.) and β-SiC nanopowder (1, 3, 5% wt.) (Alfa Aesar) were used as raw materials. The average particle size of the β-SiC nanopowder was 45-55 nm (as follows from the manufacturer's certificate). At the first stage, the required amount of β-SiC powder and/or graphite were added to isopropyl alcohol. The resulting suspension was subjected to the ultrasonic treatment using homogenizer UP200Ht for 5 min. Then, the required amount of the WC nanopowder and an additional amount of isopropyl alcohol were added to the suspension. After that, the WC + C / WC + C + SiC suspension was subjected to the ultrasonic treatment for 10 min at room temperature. The resulting composition was mixed in the planetary mill at a speed of 150 rpm for 6 h. Interruptions of 1 min/h with a simultaneous reversal of rotation were carried out to prevent excessive heating. The resulting suspension was placed in a vacuum drying chamber VD 23 and dried at a temperature of 50 °C and a pressure of 100 mbar for 2 h. The compositions were mixed in Pulverisette 6 planetary mill at room temperature in 45-60 ml of isopropyl alcohol. Milling pots and balls (⌀ 1.6 mm, 150 g) made of WC–Co were used to reduce the grinding yield. The weight of each powder batch totaled 50 g.

### 2.2. Sintering of powders

Three groups of materials were investigated: WC ceramics (Group I), WC + C graphite-added ceramics (Group II), and WC + C + SiC ceramics (Group III).

Ceramics sintering was carried out in two modes: Mode I - heating at different speeds up to the shrinkage end temperature with minimal holding time (no more than 3 min); Mode II - heating at a given speed up to the shrinkage end temperature and holding at this temperature for up to 30 min.

Group I ceramics were obtained by the SPS of the WC powder at different conditions. A continuous heating mode at a rate of 10 and 100 °C/min was applied until the end temperature of the effective shrinkage of the powder ($T_s$). Sintering took place under a uniaxial compressive stress of 70 MPa applied simultaneously with the heating start. The holding time ($t_s$) at $T_s$ was 0 and 3 min. Degassing at 950 °C for 15 min and powder preliminary pressing under 350 MPa were optionally applied. Preliminary pressing of the powder was carried out in a steel mold with an inner diameter of 12 mm using a hydraulic press Sorokin (Sorokin, Russia). The sintering modes of the WC powder are presented in detail in Table 1 and in Fig. 1a.

**Table 1. Sintering modes and characteristics of the Group I ceramic samples**

| Ceramics No. | 1 | 2 | 3 | 4 | 5 | 6 |
|---|---|---|---|---|---|---|
| **SPS modes** | | | | | | |
| Stress during preliminary pressing, MPa | 0 | 0 | 0 | 350 | 0 | 350 |
| Applied stress, MPa | 70 | 70 | 70 | 70 | 70 | 70 |
| Heating rate, °C/min | 100 | 100 | 100 | 100 | 10 | 10 |
| Holding time at 950°C, min | 0 | 0 | 15 | 15 | 15 | 15 |
| Sintering temperature $T_s$, °C | 1520 | 1520 | 1520 | 1520 | 1440 | 1440 |
| Holding time at $T_s$, min | 0 | 3 | 3 | 3 | 0 | 0 |
| **Ceramics characteristics** | | | | | | |
| $W_2C$ content, % wt. | 7.4 | 7.0 | 7.5 | 7.8 | 7.7 | 8.3 |
| Absolute density $\rho$, g/cm$^3$ | 15.65 | 15.71 | 15.71 | 15.71 | 15.50 | 15.55 |
| Relative density $\rho/\rho_{th}$, % | 98.59 | 99.06 | 98.97 | 98.98 | 97.69 | 97.93 |
| Grain size $d$, μm | 0.1 | 0.1 | 0.1 | 0.1 | 0.1 | 0.1 |
| Microhardness $HV$, GPa | 28.7 | 28.0 | 27.9 | 27.4 | 28.0 | 27.4 |
| Min. fracture toughness $K_{IFR} = K_{IC}$, MPa·m$^{1/2}$ | 5.0 | 4.9 | 4.4 | 4.9 | 4.2 | 4.3 |

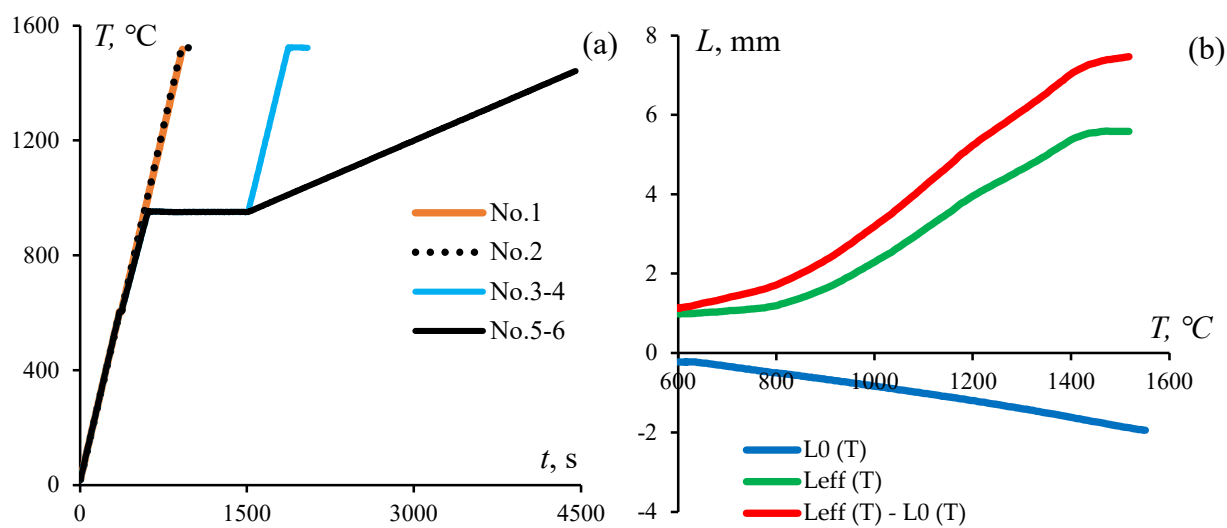

**Figure 1.** Sintering diagrams of ceramics: (a) heating diagram of the WC powder in SPS modes; (b) an approach to processing temperature curves of shrinkage of powders

SPS of WC + C and WC + C + SiC powder were conducted in a continuous heating mode at a heating rate of 50 °C/min until the end temperature ($T_s$) of the effective shrinkage of the powders. The $t_s$ was 0, 3 and 30 min. Free cooling was used to cool the samples to room temperature.

The cylindrical ceramics of Groups I–III with 12 mm in diameter and 5-6 mm high were sintered using Dr. Sinter model SPS-625 setup. Sintering was carried out in a vacuum (6 Pa) in graphite molds with an inner diameter of 12.8 mm. Graphite foil placed between the samples and the inner surface of the mold was used to reduce the clearance between the sample and the inner surface of the mold. The temperature was controlled by an optical pyrometer IR-AHS2 focused on the graphite mold surface. The accuracy of temperature measurements was ±20°C. The dilatometer as part of Dr. Sinter model SPS-625 setup was used to determine the dependencies of the effective powder shrinkage on the heating temperature $L_{eff}(T)$. Experiments on heating empty molds were performed to account for the contribution of the mold thermal expansion $L_0(T)$. The true shrinkage ($L$) was determined by the formula: $L(T) = L_{eff}(T) – L_0(T)$ (Fig.1b) (see [26]).

The kinetics of the initial sintering stage of WC + C and WC + C + SiC powder compositions was analyzed using a high-temperature dilatometer L75VS1750C. Dilatometry studies were carried out in an argon atmosphere with a purge rate of 25 ml/min under a minimum stress of the lower punch

(500 mN). The heating rate was 10°C/min. The samples were heated up to 1650°C using MoSi$_2$ heaters. The temperature was controlled using a B-type (platinum-rhodium) thermocouple.

*2.3. Characterization of powders and ceramics*

The average particle size of the WC powder ($R$) was estimated by the formula: $R = 6/(\rho_{th} \cdot S_{sp})$, where $\rho_{th}$ = 15.44 g/cm$^3$ is the theoretical density of α-WC, $S_{sp}$ is the specific surface area of the powder particles obtained by the BET method using TriStar 3000 analyzer. The content of carbon (6.26% wt.) and oxygen (0.26% wt.) in the raw WC powder was determined using an analyzer Leco CS 400.

The microstructure of the powders and ceramics was studied by scanning electron microscope (SEM) TESCAN Vega 2 and JSM-6490 with an energy dispersive (EDS) microanalyzer INCA 350 and on a transmission electron microscope (TEM) JEM-2100F. X-ray diffraction (XRD) studies were carried out on the XRD-7000 diffractometer in CuKα radiation (λ = 1.54 Å). Scanning was performed in the Bragg–Brentano geometry in the range of angles 2θ = 30–80° with the step of 0.04° and the exposure of 2 s. The Rietveld method was used to quantitative phase analysis [27]. Refinement of the experimental XRD patterns was carried out in the software DIFFRAC.TOPAS using cif-files of the ICSD (2015): α-WC (ICSD 43380), W$_2$C (ICSD 159904). The accuracy of determining the phase composition was ± 0.5% wt.

The Vickers hardness ($HV$) of the ceramics was measured with a QNESS 60 A+ EVO hardness tester. The load totaled 10 kg. The minimum fracture toughness ($K_{IC}$) was calculated based on the Palmquist model by the longest radial fracture. The elastic modulus of the ceramics was taken as $E$ = 710 GPa. The average inaccuracy of $HV$ and $K_{IC}$ was ± 0.2 GPa and ± 0.2 MPa·m$^{1/2}$, respectively. It is worth noting that the Palmquist fracture toughness does not fully reflect the physical meaning of the $K_{IC}$ fracture toughness [28]. In this regard, some authors classify Palmquist crack resistance as a group of Indentation Fracture resistance ($K_{IFR}$) techniques (see [29]). The group of Indentation Fracture Resistance (*IFR*) methods includes methods for calculating fracture toughness according to

Anstis, Niihara, Palmquist, etc. All these methods are actively used to evaluate the mechanical properties of tungsten carbide-based ceramics [30]. In denoting Palmquist fracture toughness ($K_{IFR}$) as $K_{IC}$, we followed the already established tradition in ceramic materials science [28].

The density of the samples ($\rho$) was measured by hydrostatic weighing using CPA225D scales. The theoretical density of α-WC was taken as $\rho_{th}$ = 15.77 g/cm$^3$, W$_2$C – 17.15 g/cm$^3$, SiC – 3.21 g/cm$^3$. The inaccuracy of the relative density ($\rho/\rho_{th}$) was ± 0.2%, while that of the $\rho$ was 0.05 g/cm$^3$.

The samples were mechanically polished to eliminate a carburized layer ~300-350 μm in thickness prior to investigating the microstructure, phase composition and mechanical properties. This layer is formed on the surface of the samples as a result of carbon diffusion from the graphite mold [25].

## 3. Results

### 3.1. Characteristics of the initial powders

Figures 2a and 2b show TEM images of the plasma chemical powder of tungsten carbide. The particles of the WC powder form agglomerates of a few microns in size, which consist of individual nanoparticles of ~100 nm (Fig. 2a). The average particle size of α-WC measured by the BET method is 95 nm. The results of the TEM investigations (Fig. 2b) testify that the WC powder has a crystalline structure. Twins are observable in the WC powder particles (Fig. 2b). A thin amorphous layer is also present on the surface of the WC powder particles (Fig. 2b). It is assumed that the amorphous layer consists of tungsten oxide that can cause stronger agglomeration of WC nanopowders [31,32].

Figures 2c and 2d show SEM images of the WC + C+ SiC powder. Homogeneous distribution of the SiC particles (dark gray in Fig. 2c and 2d) was ensured by the mixing.

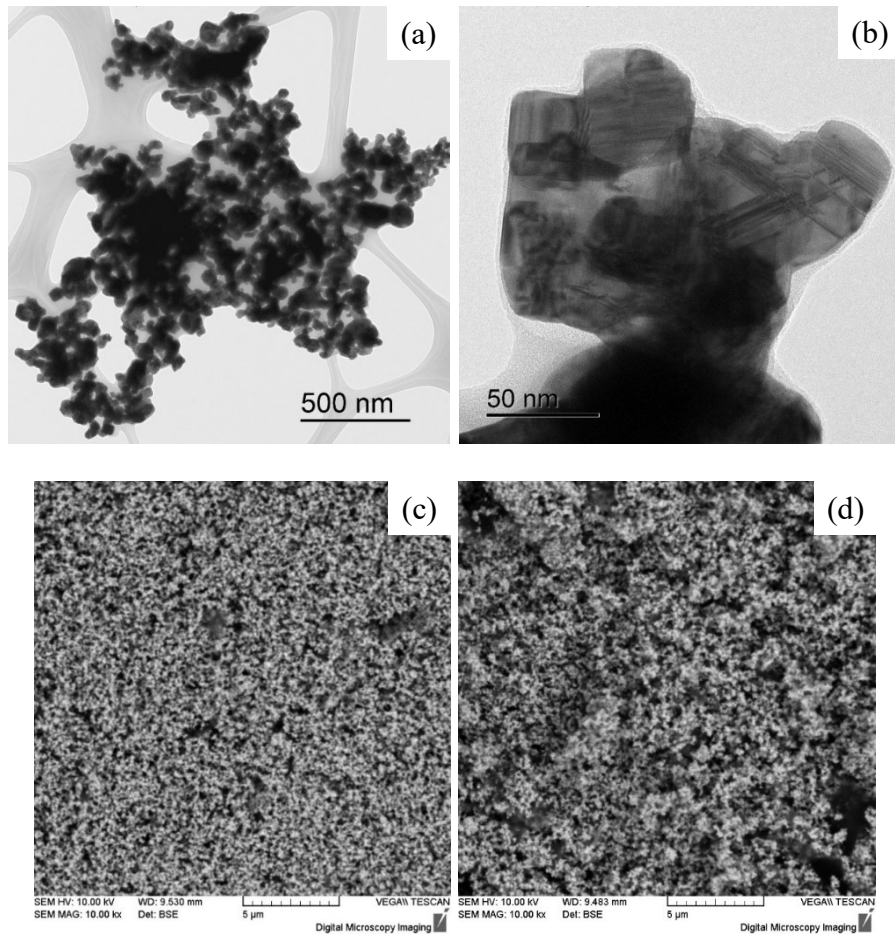

**Figure 2.** TEM images of the WC (a, b), SEM images of the WC + 3% SiC (c) and WC + 5% SiC (d) powder compositions.

Figure 3 presents the results of the XRD studies of the WC, WC + C, WC + C + SiC powders. The XRD peaks on powder patterns corresponded to α-WC phase (PDF No. 00-051-0939). No XRD peaks corresponding to other phases were detected. The XRD peaks of the WC phase were significantly broadened due to nano crystalline size of the one. No XRD peaks corresponding to β-SiC were detected (Fig. 3). This is attributed to the fact that the corundum number for the most intense XRD peak (111) of the β-SiC phase (2θ ≈ 35.6°) is 2.5 and the XRD peak (111) β-SiC overlaps with one of the most intense α-WC peaks (100) (2θ ≈ 36.6°). With the β-SiC silicon carbide content (PDF No. 00-029-1129) being low, the intensity of its peaks in XRD patterns goes down to the background level.

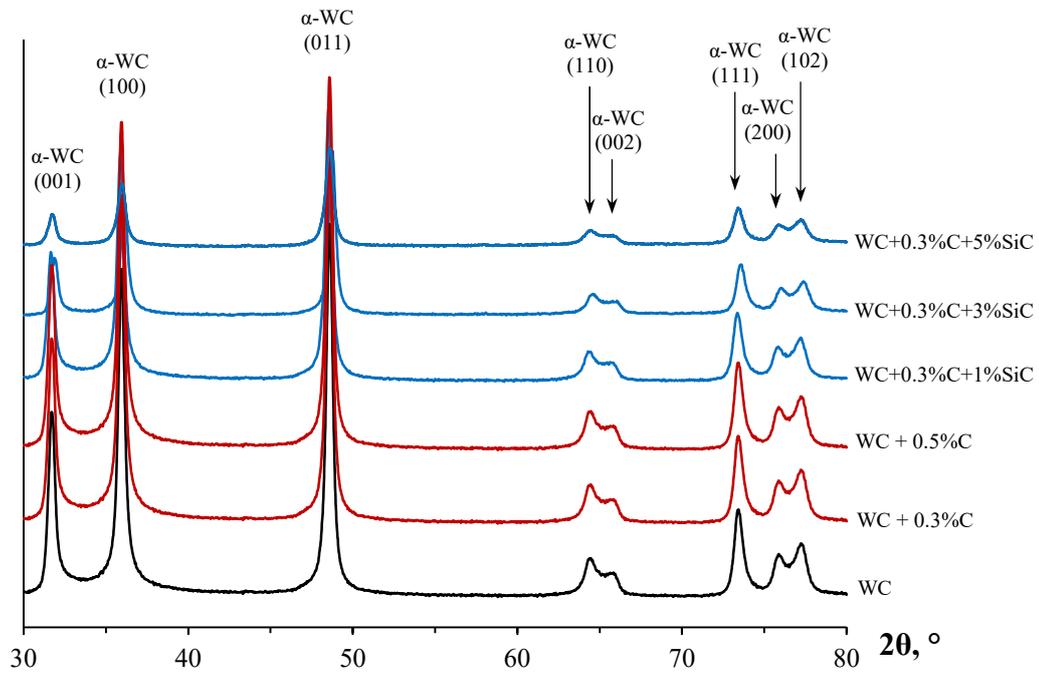

**Figure 3.** XRD patterns of the WC powder, WC + C and WC+ C + SiC powder compositions

### 3.2. Spark plasma sintering of the WC nanopowder (Group I)

SPS modes of the WC powder are presented in detail in Table 1. Figure 4 shows the SPS temperature curves for shrinkage $L(T)$ and shrinkage rate $S(T)$ of the WC powder.

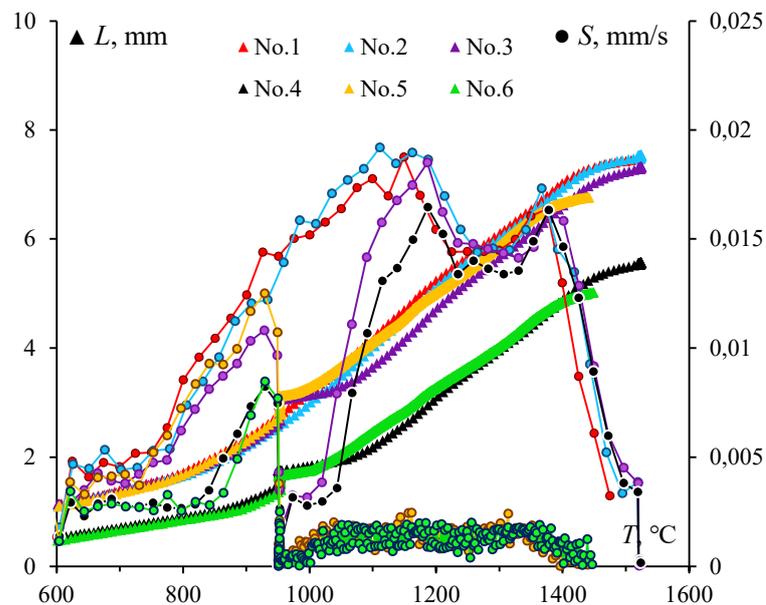

**Figure 4.** Temperature shrinkage $L(T)$ and shrinkage rate $S(T)$ curves of the WC powder at different SPS conditions (Table 1)

The $L(T)$ curves had a conventional three-stage character [14, 16, 19, 26]: slight compaction at the initial stage at below 950-1000 °C (stage I), intensive shrinkage at 1000–1400 °C (stage II), low-intensity shrinkage at above 1450-1500 °C. The maximum shrinkage value ($L_{max}$) is ~7 mm for heating modes 1, 2, 3 and 5. The $L_{max}$ value decreases to ~5–5.5 mm as a result of preliminary pressing at a pressure of 350 MPa in modes 4 and 6 (Table 1).

The $S(T)$ curves are characterized by two maxima at 1100–1200 and 1350–1400 °C in modes 1–4. The maximum shrinkage rate ($S_{max}$) for these characteristic temperatures are close to $2 \cdot 10^{-2}$ and $1.6 \cdot 10^{-2}$ mm/s, respectively. The shrinkage rate is reduced to the minimum value (~$2 \cdot 10^{-3}$ mm/s) for modes 5 and 6 because of a decrease in the heating rate from 100 °C/min (for modes 1–4) to 10 °C/min and correspondence increasing the sintering duration.

Table 1 shows that the ceramics No.2 characterized the highest density in Group I. Reducing the heating rate from 100 to 10 °C/min led to a decrease in the shrinkage end temperature from 1520 to 1440 °C (at $t_s$ = 0 min) and the relative density of the ceramics reduces from 98.59 to 97.69%. Increasing the holding time to $t_s$ = 3 min led to an increase in the density of the ceramics from 98.59 to 99.06%. Degassing at 950 °C (for 15 min) has no significant effect on the density of the samples. The density of the ceramics sintered from the preliminary pressed powder within the measurement error (± 0.03%) coincides with the density of the ceramics sintered from the powder that has not been preliminary pressed.

Fig. 5 presents the microstructure of the Group I ceramic samples sintered under different SPS modes. Variation of the SPS mode led to no a significant change in the microstructure of the samples. The ceramics have a homogeneous UFG microstructure with an average grain size of ~100–200 nm (Fig. 6). Some samples show meso- and microstructural heterogeneity (Fig. 5d, e). The microstructure heterogeneity observed is probably due to the agglomeration in the initial WC powder.

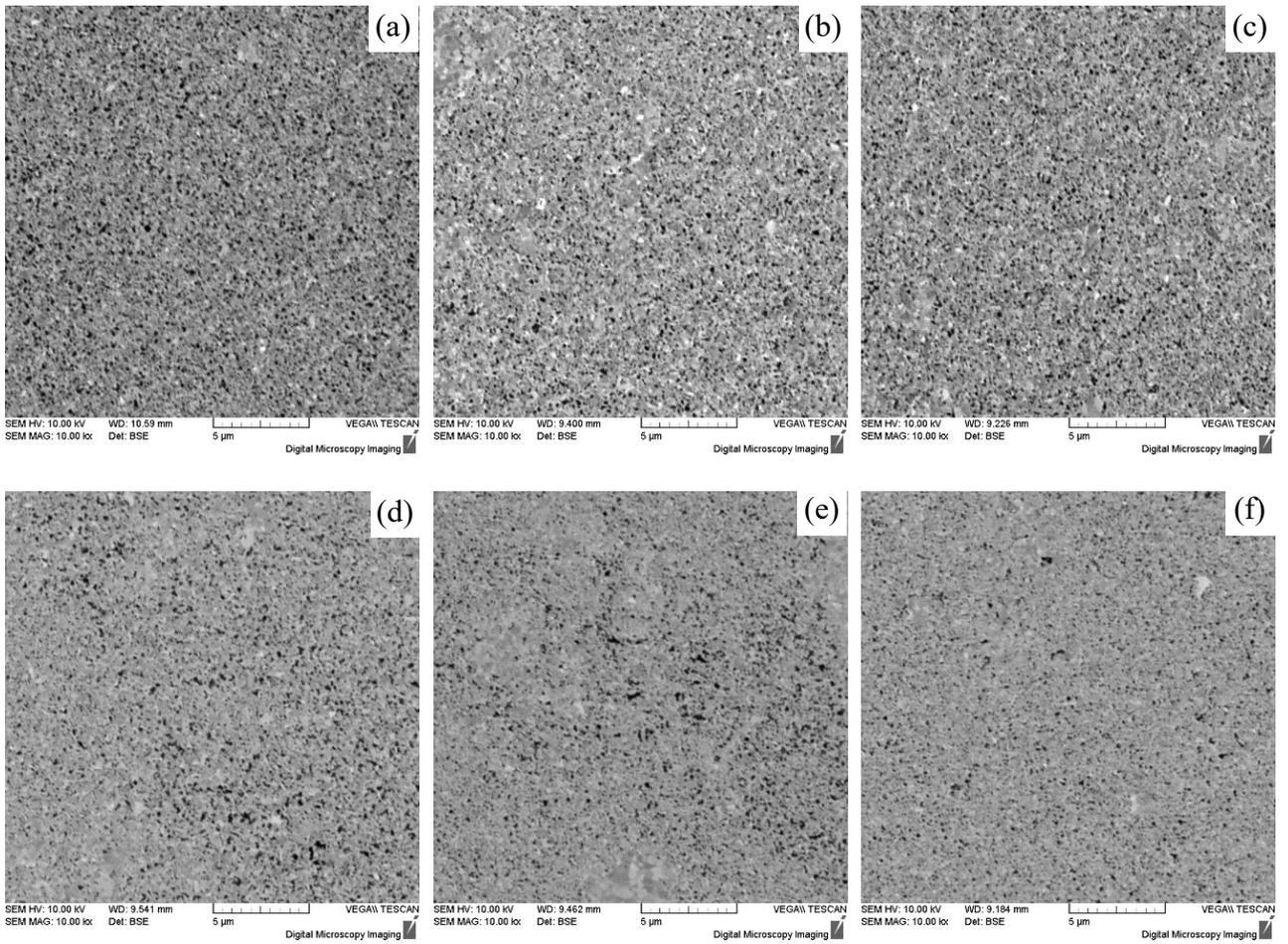

**Figure 5.** Microstructure of Group I ceramics No.1 (a), No.2 (b), No.3 (c), No.4 (d), No.5 (e), No.6 (f) sintered using different SPS modes (Table 1). SEM

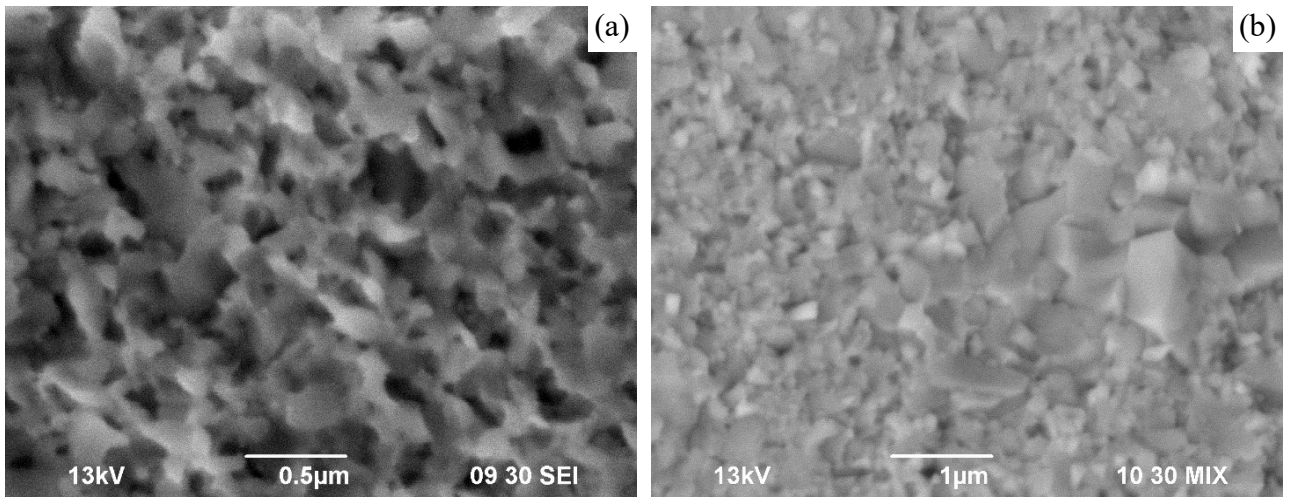

**Figure 6.** Microstructure of Group I ceramics No.2 (see Table 1): (a) electropolishing; (b) sample fracture. SEM

Fig. 7 presents the XRD patterns of the Group I ceramic. The XRD peaks corresponded to α-WC (PDF No. 00-051-0939) and $W_2C$ (PDF No. 00-035-0776) are observed on powder patterns. Content of $W_2C$ phase in the Group I samples is presented in Table 1. The minimum content of $W_2C$ is observed in ceramic No.2. Degassing at 950 °C leads to an increase in the $W_2C$ content up to ~8% wt. The adsorbed oxygen interaction with carbon during degassing leads to a decrease in the carbon content in the sintered volume and, consequently, an increase in the $W_2C$ content [31, 32]. Reducing the heating rate from 100 to 10 °C/min increased the $W_2C$ content from 7.0 to 7.7–8.3% wt.

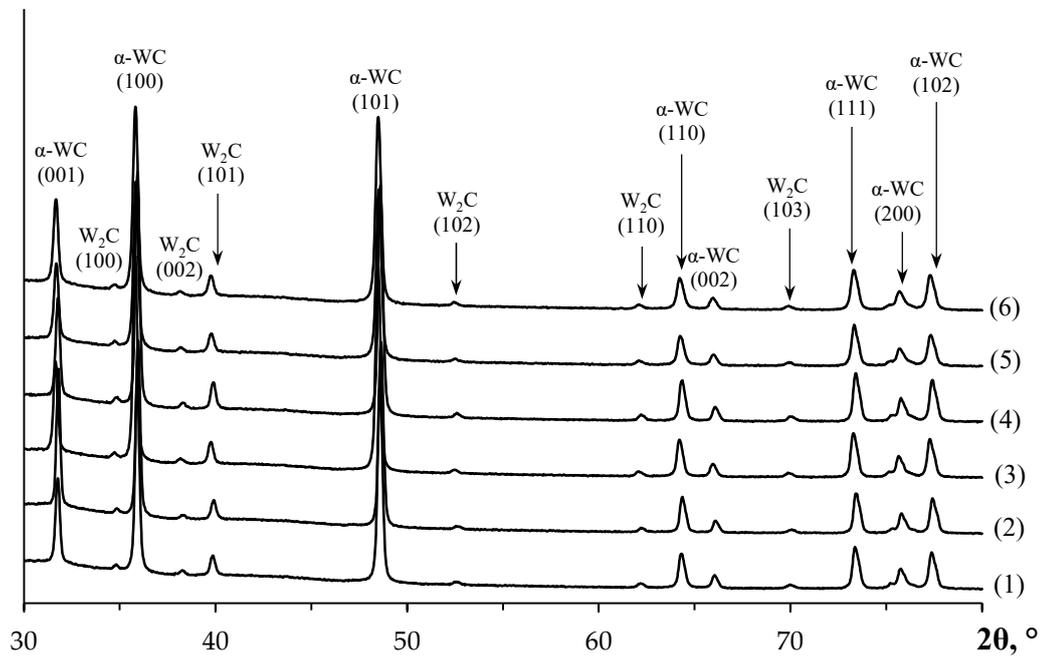

**Figure 7.** XRD patterns of the Group I ceramics

It can be seen from the data in Table 1 that the best combination of hardness and fracture toughness (FIR) was obtained in ceramic No.1: $HV$ = 28.7 GPa, $K_{IC}$ = 5.0 MPa·m$^{1/2}$. Degassing and a decrease in the heating rate from 100 to 10 °C/min reduce hardness to 27.4–27.9 GPa and fracture toughness (Indentation Fracture resistance) to 4.2–4.3 MPa·m$^{1/2}$, respectively. Cracks spread in a straight line, no significant deviation of cracks from the rectilinear trajectory was revealed (Fig. 8).

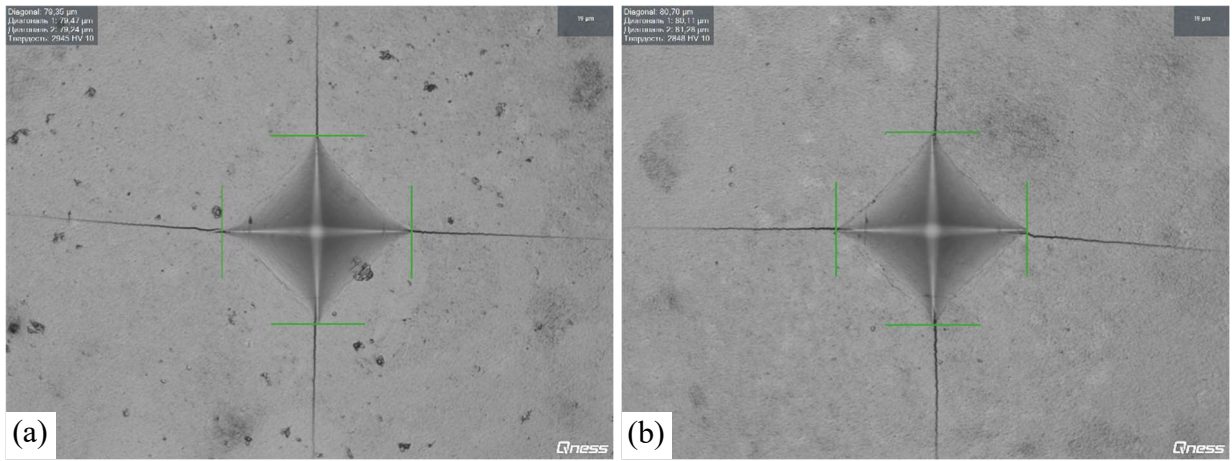

**Figure 8.** Images of the indenter imprint and crack propagation on ceramics No.1 (a) and No. 3 (b)

*3.3. Spark plasma sintering of the WC + C nanopowders (Group II)*

SPS modes of the WC + C powder compositions are presented in detail in Table 2.

**Table 2.** Sintering modes differences and characteristics of the Group II (WC + C) ceramic.

| C, % | $t_s$, min | $T_s$, °C | $\rho/\rho_{th}$, % (±0.05%) | d, μm | $W_2C$, % | HV, GPa | $K_{IFR} = K_{IC}$, MPa·m$^{1/2}$ | SPS activation energy | |
|---|---|---|---|---|---|---|---|---|---|
| | | | | | | | | $mQ_{sII}$, kT$_m$ (kJ/mol) | $Q_{sIII}$, kT$_m$ (kJ/mol) |
| 0 | 0 | 1500 | 99.4 | 0.1 | 7.5 | 27.8 | 4.0 | 1.3 (33) | 11.0 (279) |
| | 3 | | 99.7 | 0.15 | | 28.2 | 3.9 | | |
| | 30 | | 99.6 | 3.0 | | 25.3 | 5.0 | | |
| 0.3 | 0 | 1500 | 98.5 | 0.1 | 2.5 | 28.8 | 3.6 | 2.0 (51) | 13.0 (330) |
| | 3 | | 99.1 | 0.2 | | 27.6 | 4.1 | | |
| | 30 | | 98.6 | 5.5 | | 24.1 | 5.2 | | |
| 0.5 | 0 | 1400 | 97.7 | 22 | 0 | 12.9 | 9.0 | 2.3 (58) | 12.0 (304) |
| | 3 | | 97.7 | 85 | | 12.3 | 8. | | |
| | 30 | | 97.9 | 200 | | 11.7 | 8.5 | | |

Fig. 9 shows the temperature curves $L(T)$ and $S(T)$ obtained by the dilatometry studies (Fig. 9a) and by the SPS (Fig. 9b) of the WC + C nanopowders. The shrinkage and shrinkage rate curves for the WC + 0.5% C nanopowder differ from the ones for the other powders (Fig. 9a). The maximum shrinkage $L_{max}$ of the WC + 0.5% C composition is significantly higher than that of the WC and WC + 0.3% C one, despite the same initial powder mass. The $S(T)$ curve for the WC + 0.5%C nanopowder peaks at ~950 °C, which is associated with degassing, another peak occurs at ~1620 °C. There are three peaks on the $S(T)$ curves of WC + 0.1%C and WC + 0.3%C powders under study: the first peak at 960–1000 °C; the second peak at 1400 °C; the beginning of the third peak at 1650 °C.

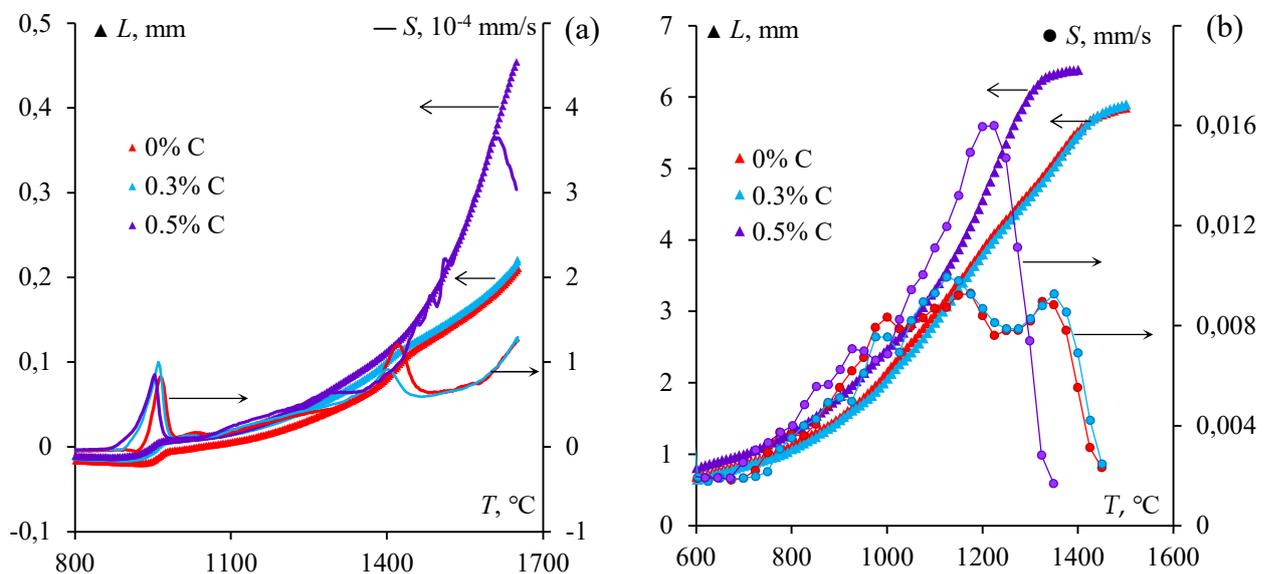

**Figure 9.** Temperature shrinkage and shrinkage rate curves of the WC powders with varying graphite content: dilatometry studies of CPS (a) and SPS (b)

Figure 9b shows the temperature curves $L(T)$ and $S(T)$ of the WC + C powders during SPS. A conventional three-stage sintering character can be observed. The addition of 0.3% C into the α-WC nanopowder leads to no noticeable change in its shrinkage kinetics, while the addition of 0.5% C causes a shift of the $L(T)$ curve to lower temperatures (Fig. 9). The temperature of the intensive shrinkage end (Stage II) decreases by 100 °C and reaches 1400 °C. The value of $L_{max}$ increases from 5.9 to 6.4 mm with the graphite concentration growing from 0.3 to 0.5%, while $S_{max}$ increases from $1.5 \cdot 10^{-2}$ to $2.6 \cdot 10^2$ mm/s.

The S(T) curve for the WC powder with 0.5% graphite has a more complex character than those for the WC and WC + 0.3% C powders. In the case of the WC powders without graphite and with 0.3% C, the S(T) curves show two peaks at 1150 and 1375 °C. There is a monotonic increase in the shrinkage rate to $2.7 \cdot 10^{-2}$ mm/s at 1250 °C followed by a sharp decrease in the shrinkage rate to $1.3 \cdot 10^{-3}$ mm/s at 1375 °C during SPS of the WC + 0.5% C nanopowder.

Table 2 shows that the binderless WC ceramics have high relative density (≥ 99.5%). The addition of 0.3% and 0.5% graphite reduces the relative density of the ceramics to ~98.7% and ~97.7%, respectively.

Fig. 10 shows the SEM images of the ceramics with varying graphite content. Microstructural examination shows that an increase in the graphite concentrations from 0 to 0.5% wt. leads to abnormally large grains with a length of 50-100 μm (Fig. 10a, 10d, 10g) [6, 7]. An increase in the isothermal holding time from 0 to 3 min during sintering of the WC and WC + 0.3% C samples leads to a slight increase in the average grain size by 50-100 nm (Fig. 10a, 10b, 10d, 10e). An increase in the holding time to 30 min resulted in an increase in the average grain size to 3 μm for the WC ceramics and to 5.5 μm for the WC + 0.3% C ceramics (Fig. 10c, 10f). An increase in the holding time from 0 to 30 min during sintering of the WC + 0.5% C samples leads to an increase in abnormal grain size to 200 μm (Fig. 10g, 10h). Abnormally large grains in the WC + 0.5% C samples after holding for 30 min occupy almost the entire cross section of the ceramic (Fig. 10i).

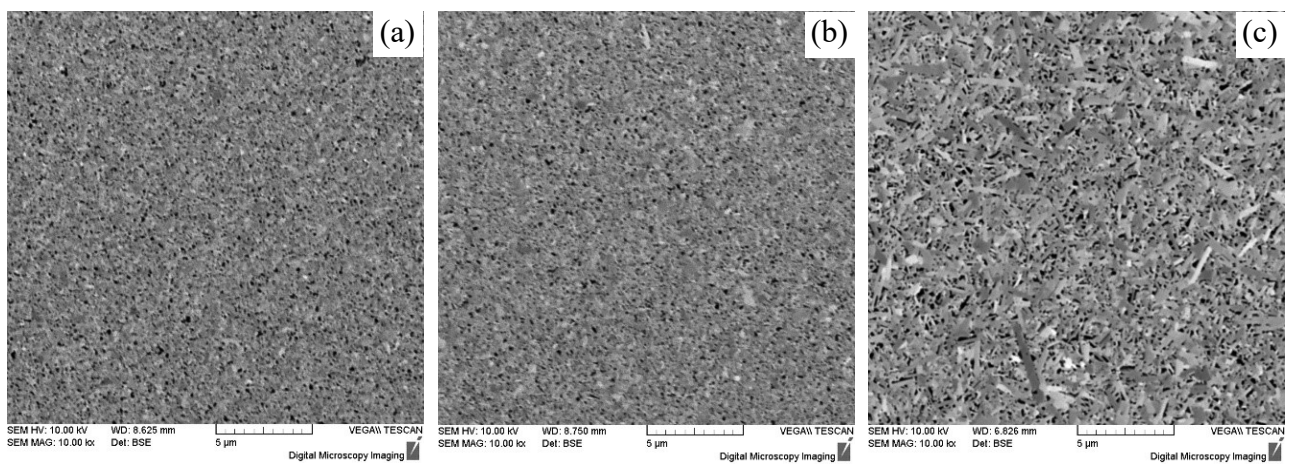

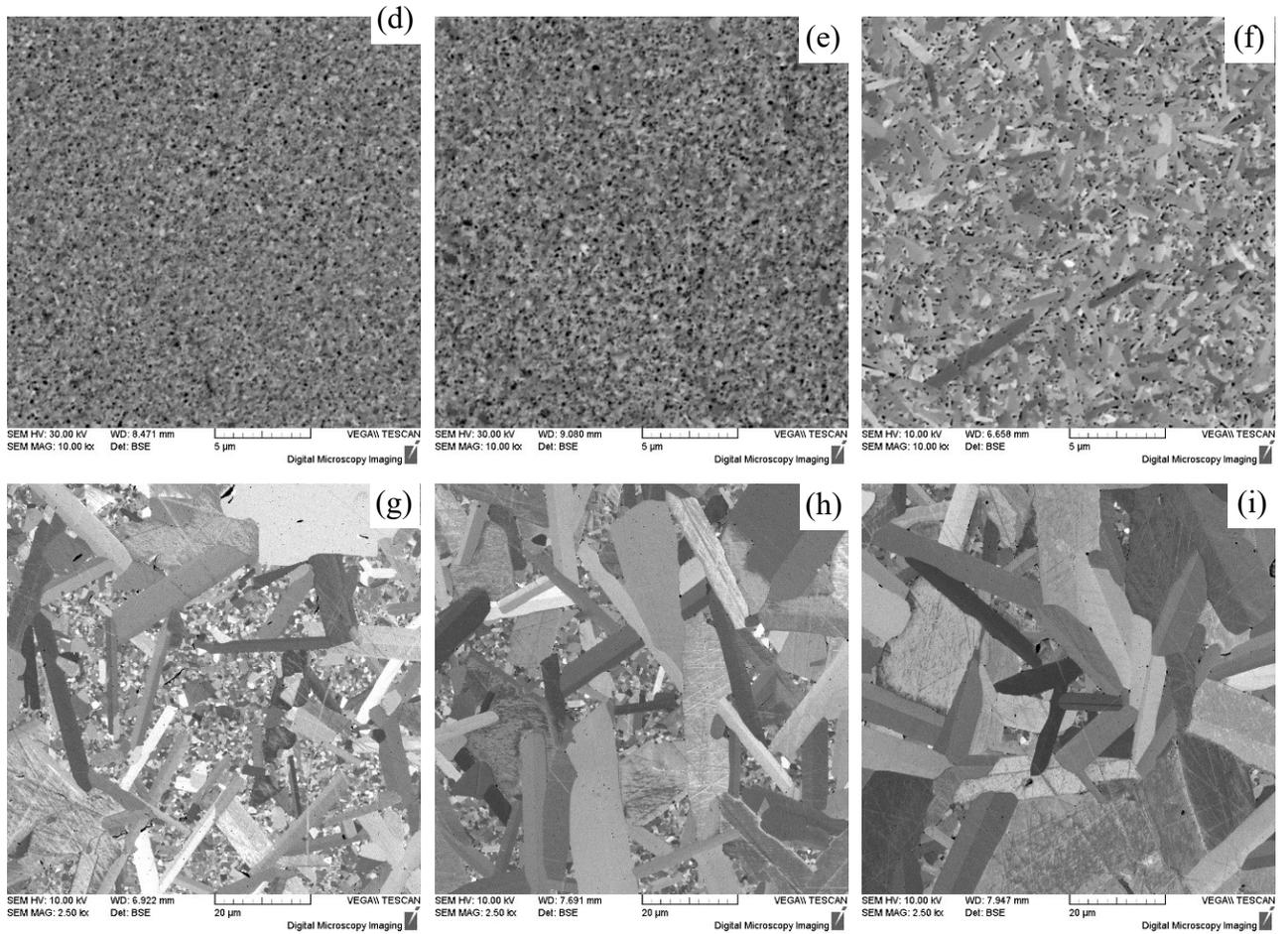

**Figure 10**. Microstructure of the WC (a, b, c), WC + 0.3% C (d, e, f) and WC + 0.5% C (g, h, i) ceramics with different holding times at $T_s$: 0 min (a, d, g), 3 min (b, e, h) and 30 min (c, f, i). SEM

The analysis of the XRD patterns shown in Fig. 11 demonstrated that the WC and WC + 0.3% C ceramics contain WC and $W_2C$ phases. The $W_2C$ phase content in the ceramics sintered from the α-WC nanopowder without adding graphite is ~7.5% wt. and does not depend on the isothermal exposure time. The fraction of the $W_2C$ particles decreases to ~2.5% wt. with an increase in the graphite concentration to 0.3% wt. There are no peaks corresponding to the $W_2C$ phase in the XRD patterns of the WC ceramic with 0.5% wt. of graphite.

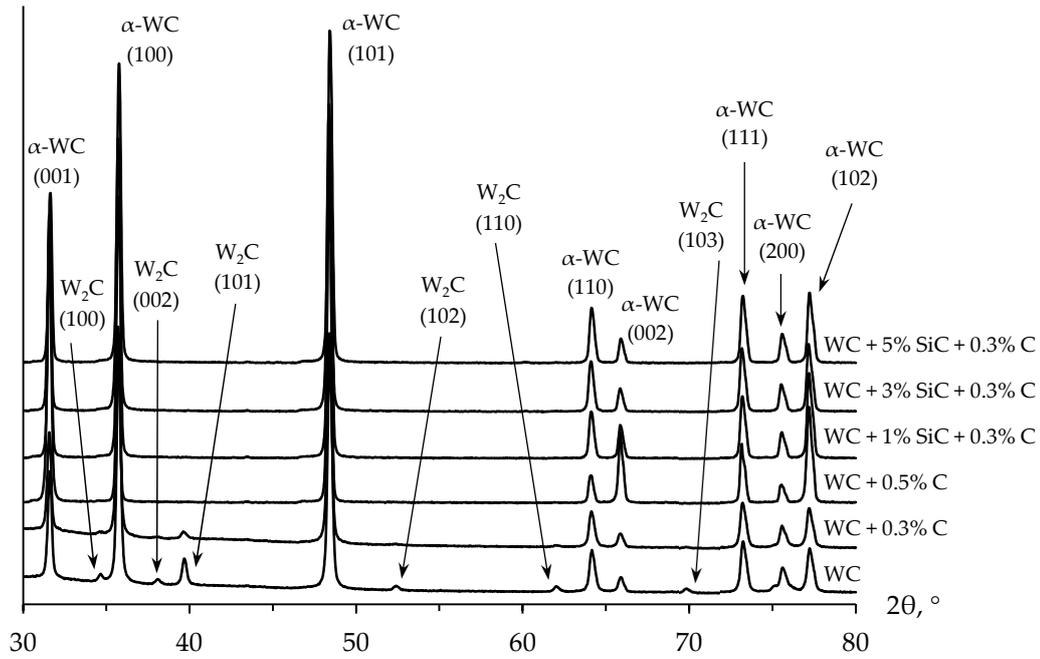

**Figure 11**. XRD patterns of the Group II (WC + C) and Group III (WC + SiC + 0.3% C) WC ceramics.

Ceramics produced of the α-WC nanopowder are characterized by hardness $HV$ = 27.8 GPa and fracture toughness $K_{IC}$ = 4 MPa·m$^{1/2}$ (see Tab. 2). An increase in isothermal holding time from 0 to 30 min reduces $HV$ to 25.3 GPa and increases $K_{IC}$ to 5 MPa·m$^{1/2}$. The addition of 0.3% wt. of graphite into the nanopowders and, consequently, a decrease in W$_2$C content, increases $H_V$ to 28.8 GPa while reducing $K_{IC}$ to 3.6 MPa·m$^{1/2}$. Ceramics with the addition of 0.5% wt. of graphite have a low hardness $HV$ = 11-12 GPa and an increased $K_{IC}$ up to 9 MPa·m$^{1/2}$. A longer isothermal holding time reduces the hardness of the WC + C ceramics and slightly increases $K_{IC}$ (see Table 2).

*3.4. Spark plasma sintering of the WC + SiC + 0.3% C nanopowder (Group III)*

SPS modes of the WC + SiC + 0.3% C powder are presented in detail in Table 3.

Fig. 12a shows the temperature curves $L(T)$ and $S(T)$ obtained by the dilatometry of the WC + SiC powders with an addition of 0.3% graphite. An increase in the SiC content leads to a slight increase in the shrinkage. The $L(T)$ curves have a complex multistage character. The $S(T)$ curves for the WC + SiC powders show two main maxima at 900 °C and 1400 °C. The maximum shrinkage rate at 900 °C is almost independent of the SiC content, while $S_{max}$ at 1400 °C increases along with an increase in the SiC content. As the temperature increases above 1500-1550 °C, the shrinkage rate of powders in dilatometric studies starts to increase again (Fig. 12a).

**Table 3.** Sintering modes differences and characteristics of the Group II (WC + SiC + 0.3% C) ceramics

| SiC, % | $t_s$, min | $T_s$, °C | $\rho/\rho_{th}$, % | $d$, µm | $W_2C$, % | $HV$, GPa | $K_{IFR} = K_{IC}$, MPa·m$^{1/2}$ | SPS activation energy | |
|---|---|---|---|---|---|---|---|---|---|
| | | | | | | | | $mQ_{sII}$, kT$_m$ (kJ/mol) | $Q_{sIII}$, kT$_m$ (kJ/mol) |
| 1 | 0 | 1350 | 97.4 | 7 | 0 | 19.2 | 5.9 | 2.2 (56) | 10.0 (250) |
| | 3 | | 97.3 | 10 | | 18.1 | 6.1 | | |
| | 30 | | 97.7 | 16 | | 17.3 | 6.2 | | |
| 3 | 0 | 1350 | 95.5 | 0.5 | 0 | 22.6 | 3.8 | 2.2 (56) | 12.0 (304) |
| | 3 | | 95.4 | 1.2 | | 21.1 | 4.1 | | |
| | 30 | | 96.7 | 12 | | 18.3 | 4.4 | | |
| 5 | 0 | 1350 | 95.1 | 0.3 | 0 | 21.5 | 3.9 | 2.15 (55) | 10.0 (250) |
| | 3 | | 96.1 | 1 | | 20.1 | 4 | | |
| | 30 | | 95.8 | 3.2 | | 18.7 | 4.4 | | |

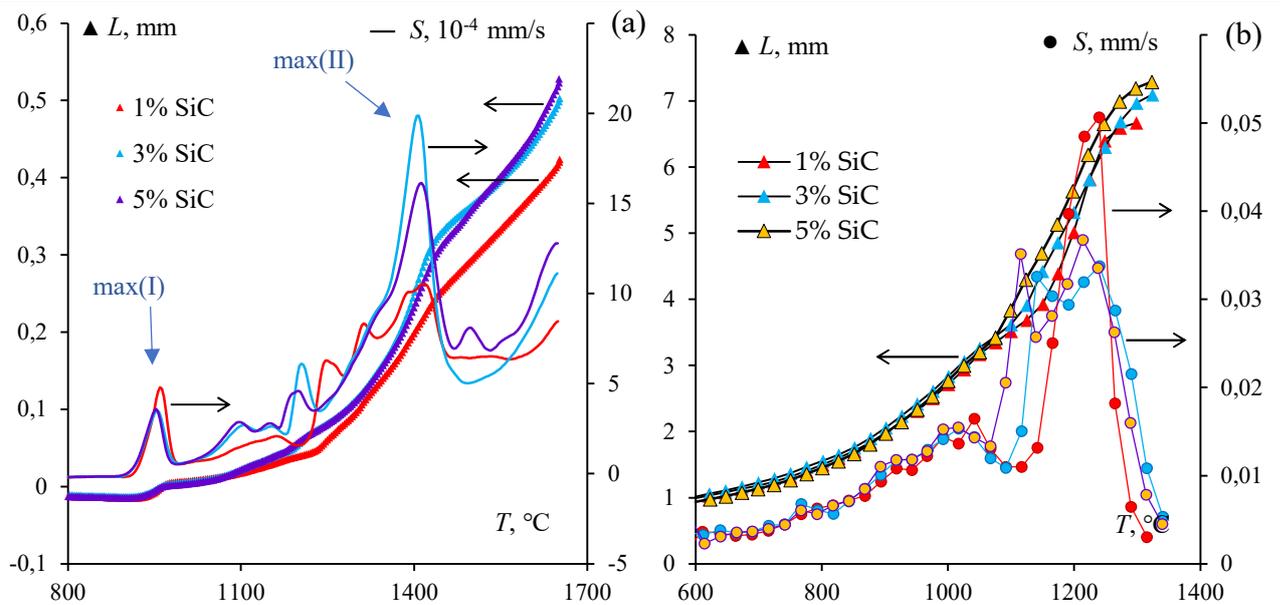

**Figure 12.** Temperature shrinkage and shrinkage rate curves of the WC + 0.3%C powders with varying SiC content: dilatometry studies of CPS (a) and SPS (b)

Fig. 12b shows the temperature curves for the shrinkage $L(T)$ and shrinkage rate $S(T)$ of the WC + 0.3% C + SiC powder compositions during SPS. An increase in the SiC content from 1 to 5% wt. does not change the shrinkage kinetics of the WC nanopowder compositions. The optimal sintering temperature for the WC + SiC + 0.3% C powders is 1350 °C, which is 150 °C below the maximum shrinkage temperature of pure WC nanopowder. The $L(T)$ curves for the WC + SiC ceramics have a conventional three-stage character, which are divided into: slight densification at low temperatures of 600–1000 °C (Stage I), intense shrinkage in the medium temperature range of 1300–1450 °C (Stage II) and, again, weak shrinkage change at 1270–1350 °C (Stage III).

The $S(T)$ curves for the WC + SiC + 0.3% C powders have a more complex character - two shrinkage rate maxima ($S_{max(1)}$, $S_{max(2)}$) are observed at ~1000°C ($S_{max(1)}$) and 1250°C ($S_{max(2)}$) (Fig. 12b). As the SiC content increases from 1 to 5%, the maximum shrinkage rate $S_{max(2)}$ increases from $10^{-2}$ to $4 \cdot 10^{-2}$ mm/s at 1250–1300°C; the maximum shrinkage $L_{max}$ increases from 5.8 to 7.3 mm (Fig. 12a).

Table 3 shows that the SPS allows to obtain the WC + SiC ceramics with high relative density. The density of binderless WC samples with 0.3% of graphite after holding for 3 min at the maximum sintering temperature is 99.1% (Table 2). An increase in the SiC content to 5% reduces the density of ceramics to 95.1%. Longer isothermal holding time raises the density of WC + SiC + 0.3%C ceramics. This effect is even more noticeable in ceramics with a higher content of SiC particles.

Fig. 13 shows the SEM images of the WC + SiC + 0.3% C ceramics microstructure. An increase in the SiC content from 1 to 5% leads to a noticeable increase in the volume fraction of black areas in the ceramics. The average size of these areas is ~0.1–0.2 μm, which is comparable in size to SiC particles that form small agglomerates (Fig. 2d). There is a marked increase in abnormally large grains in the WC + 1% SiC + 0.3% C ceramic. An increase in the SiC content to 3 and 5% reduces the share of abnormally large grains, whereas the average matrix grain size rises to 0.5 μm.

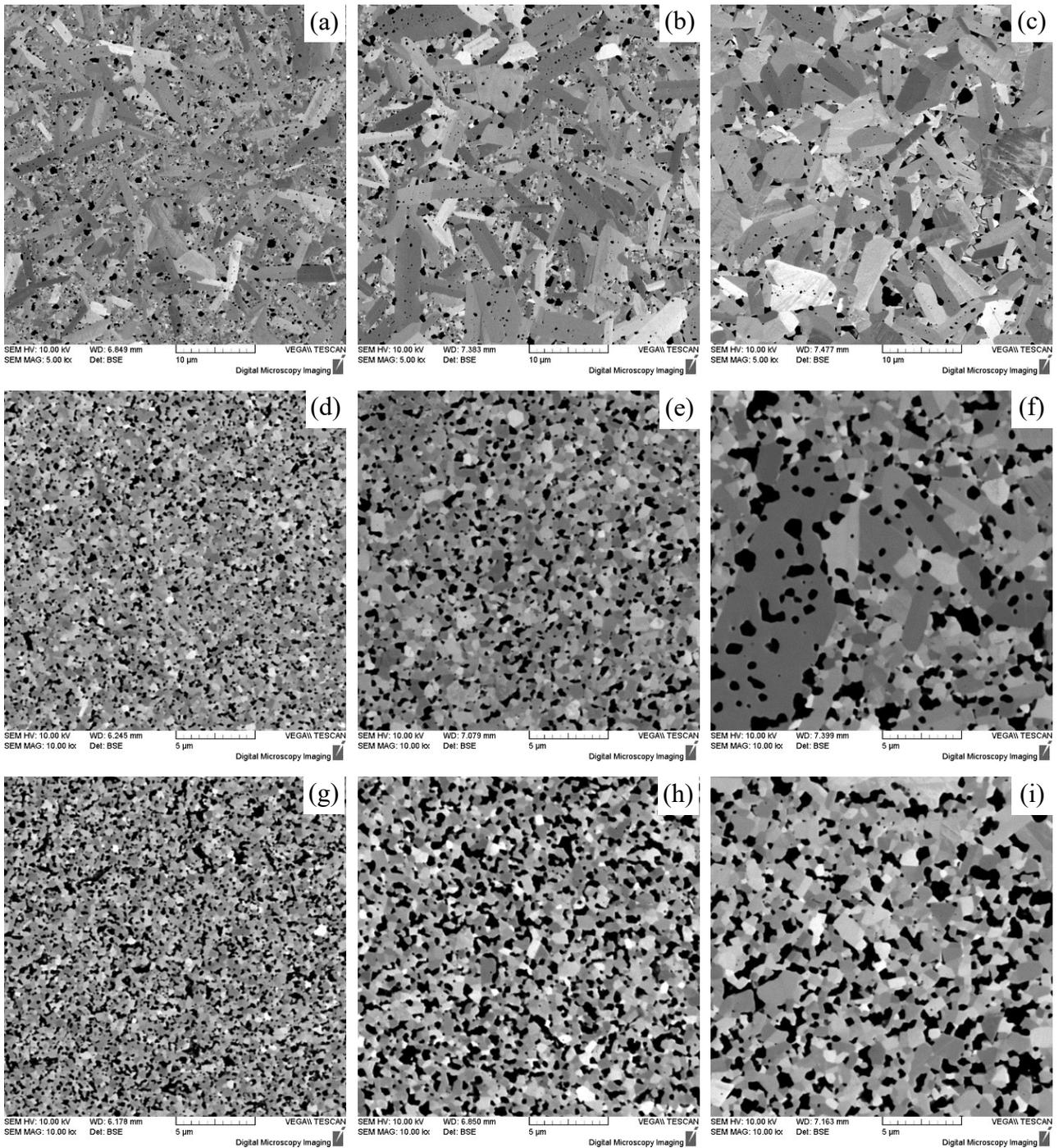

**Figure 13.** Microstructure of the WC + 0.3% C ceramics with 1% (a, b, c), 3% (d, e, f), 5% SiC (g, h, i) obtained by SPS holded for $t_s$ = 0 min (a, d, g), 3 min (b, e, h) and 30 min (c, f, i) at $T_s$ = 1350 °C. SEM

An increase in the holding time ($t_s$) at $T_s$ = 1350 °C raises the average grain size in ceramic contained 1% SiC and increases the volume fraction of abnormally large grains (Fig. 12c). It should be noted that when $t_s$ increases from 3 to 30 min in the microstructure of WC + 0.3% C and WC + 1% SiC

+ 0.3% C ceramics, the size and volume fraction of abnormal grains growing in crystallographically preferred directions are observed to increase. An increase in the isothermal holding time $t_s$ for ceramics with 3 and 5% SiC does not lead to abnormally large grains, but there is a fairly noticeable increase in the average matrix grain size (Fig. 13).

XRD patterns of Group III ceramics presented on Fig. 11 show no XRD peaks corresponding to the $W_2C$ phase. No peaks corresponding to any Si-bearing phases are observed in the XRD patterns.

WC + SiC ceramics have sufficiently high mechanical properties. WC + 1% SiC + 0.3% C samples have high hardness $HV = 19.1$ GPa, $K_{IC} = 5.2$ MPa·m$^{1/2}$ and relative density 97.4%. A further increase in the content of the SiC particles to 3 and 5% leads to no noticeable increase in hardness, but contributes to a slight decrease in IFR.

Fig. 14 shows photographs of cracks formed when $K_{IFR} = K_{IC}$ was measured using the Vickers pyramid. Images in Fig. 14 show that fracture tips have a branching character, meaning that fractures are predominantly distributed along grain boundaries. Comparing Figs. 14 and 15, we can conclude that the appearance of a certain proportion of abnormally large grains leads to a deviation of the crack propagation trajectory from the straight-line trajectory. Increasing the holding time leads to the formation of a uniform coarse-grained ceramic microstructure and to the appearance of large rectilinear cracks. In ceramics with increased SiC content, the character of crack propagation depends on the isothermal holding time. At low isothermal holding times, anomalous grain growth leads to the formation of bimodal microstructure and, consequently, to the appearance of branching cracks (Fig. 14c, Fig. 15a, b). Increasing the holding time necessary to achieve high density leads to the formation of coarse-grained microstructure and the appearance of large rectilinear cracks despite the high content of SiC particles (Fig. 14i). With increasing size and volume fraction of anomalously large grains, areas of transcrystalline crack propagation are observed (Fig. 15c). At higher graphite content and increasing isothermal holding time, the size of anomalous grains becomes so large that cracks cannot circumvent such large grains (Fig. 15d). In this case, an increase in the length of crack sections propagating inside the WC grains is observed (Fig. 15d).

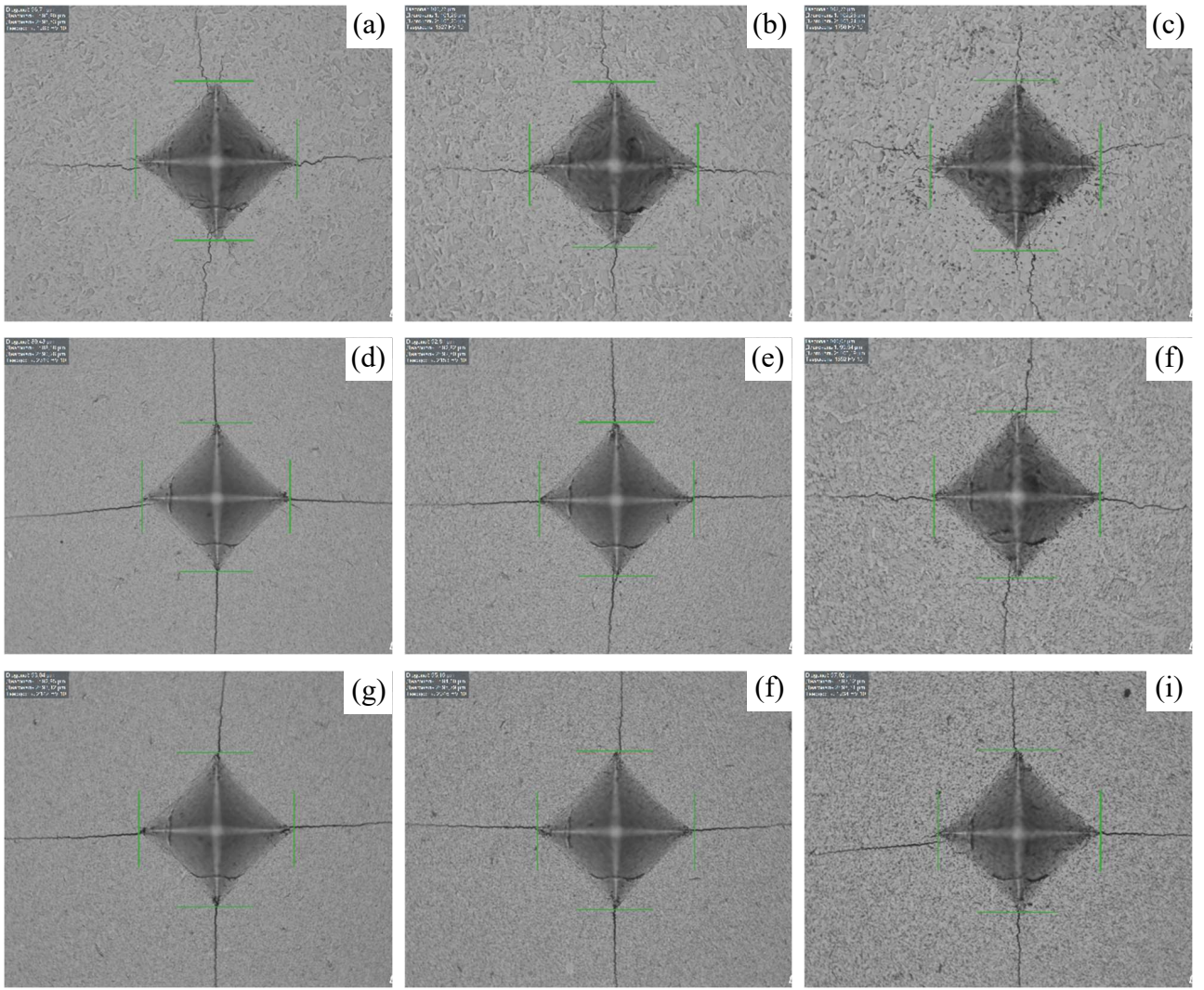

**Figure 14.** Hardness indentation images during hardness and IFR tests of the WC + 0.3%C ceramics with 1% (a, b, c), 3% (d, e, f), 5%SiC (g, h, i) obtained by SPS with isothermal holding for $t_s = 0$ min (a, d, g), 3 min (b, e, h) and 30 min (c, f, i) at $T_s = 1350$ °C

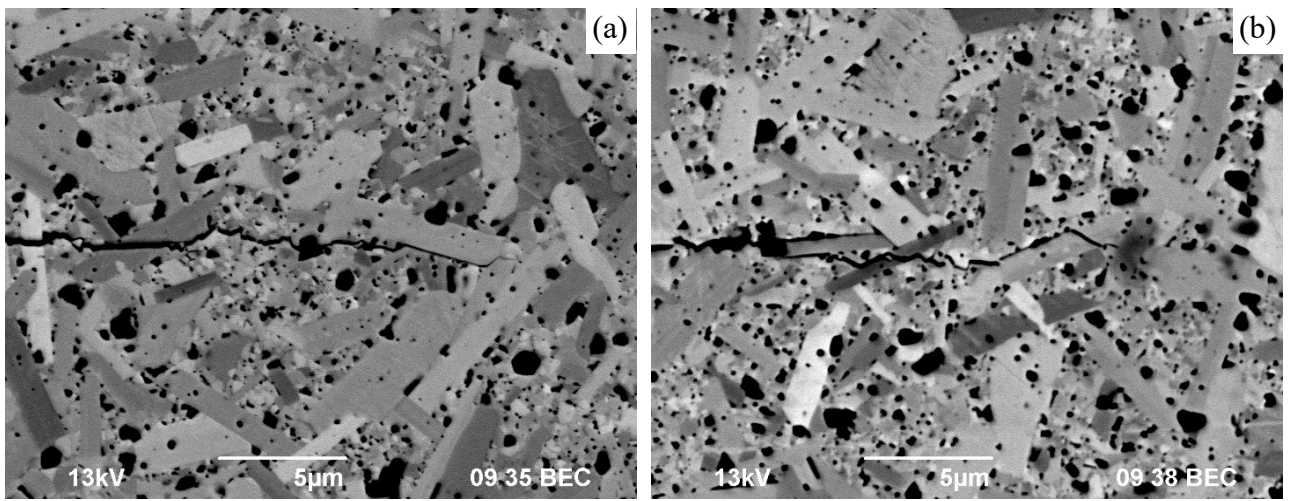

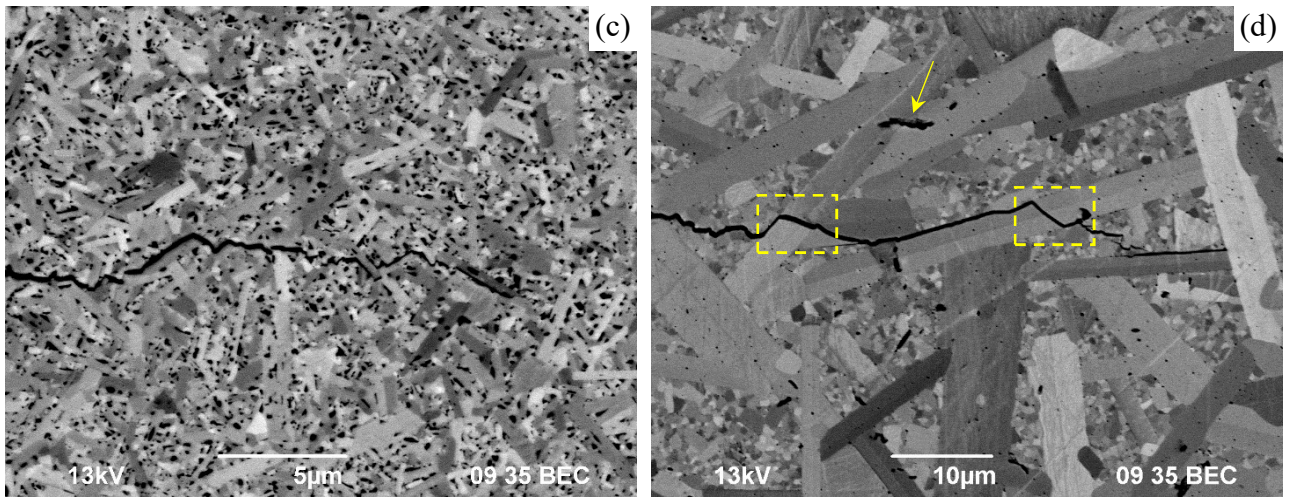

**Figure 15.** SEM images of the crack tip formed after fracture resistance measurement by Indentation Fracture Resistance test: (a, b) WC + 1%SiC + 0.4%C ceramics; (c) WC + 0.3%C ceramics; (d) WC + 0.5%C ceramics. The areas of transcrystalline crack propagation in Fig. 15d are marked with dashed lines. The arrow in Fig. 15d marks a fragment of a large graphite particle.

Thus, three factors influence the crack propagation path in WC + SiC ceramics: (i) the character of the microstructure (grain size and degree of microstructure inhomogeneity; (ii) the pore volume fraction; (iii) the SiC content. By varying the graphite and SiC content and varying the sintering conditions, a microstructure with highly branching cracks can be formed. Such microstructure will provide high crack inhibition resistance of WC + SiC + C ceramics under IFR test conditions. The authors consider it important to emphasize once again that the $K_{IFR}$ value estimated by the length of the formed crack does not fully reflect the fracture toughness of the $K_{IC}$ material (see [28,29]). The values of $K_{IC} = K_{IFR}$ presented in Tables 1-3 indicate a high level of resistance of WC-based UFG ceramics to microcrack initiation.

### 4. Discussion

As shown in the Results section, the $L(T)$ curves during SPS have a three-stage character: low-temperature Stage I – low intensity of powder shrinkage, Stage II – intensive shrinkage, high-temperature Stage III – low intensity of powder shrinkage (Fig. 4, 7). The processes occurring at each of the stages are described in detail in [33–36].

The kinetics of the nanopowder compositions shrinking at the intensive compaction stage (Stage II) can be analyzed using the Young-Cutler model [37]. This model describes the initial stage of non-isothermal sintering of spherical particles accompanied by simultaneous processes of volume and grain boundary diffusion as well as plastic deformation:

$$\varepsilon^2(\partial\varepsilon/\partial t) = (2{,}63\gamma\Omega D_v\varepsilon/kTd^3) + (0{,}7\gamma\Omega bD_b/kTd^4) + (Ap\varepsilon^2 D/kT), \quad (1)$$

where $\varepsilon$ – relative shrinkage, $t$ – time, $\gamma$ – free energy, $D_v$ – volume diffusion coefficient, $D_b$ – grain boundary diffusion coefficient, $d$ – grain size, $p$ – stress, $D$ – diffusion coefficient during plastic deformation. According to [37,38], the slope of $\ln(T\partial\varepsilon/\partial T) - T_m/T$ dependence corresponds to the effective activation energy of the sintering process ($mQ_{sII}$), where m is a numerical coefficient that depends on the diffusion mechanism, $T_m = 3053$ K is the melting point of α-WC. According to [37,38], $m = 1/3$ for grain boundary diffusion, $m = ½$ for volume diffusion, $m = 1$ for viscous flow (creep).

Fig. 16 show $\ln(T\partial\varepsilon/\partial T) - T_m/T$ dependences for the WC ceramics with varying graphite content (Fig. 16a) and WC + SiC + 0.3% C ceramics (Fig. 16b) produced by SPS. The curves for all ceramics have a conventional two-stage character, which indirectly testifies to the validity of the Young-Cutler model [34]. The correctness of applying this approach in the analysis of the $L(T)$ curves was also demonstrated in the result analysis of SPS of WC and WC–Co powders [25, 39].

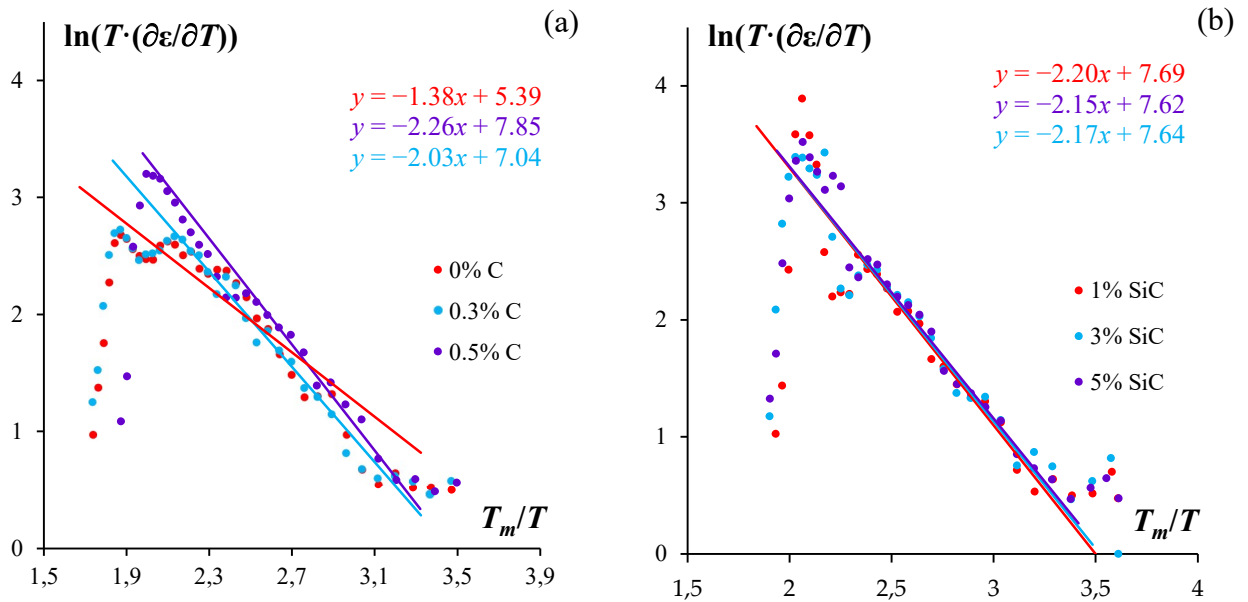

**Figure 16.** Temperature curves $L(T)$ for WC + C (a) and WC + SiC (b) powders in $\ln(T\partial\varepsilon/\partial T) - T_m/T$ coordinates. SPS activation energy calculation for Stage II

The average values of the effective SPS activation energy for Group II and Group III ceramics at Stage II ($mQ_{sII}$) are presented in Tables 2 and 3. The accuracy of the effective activation energy $mQ_{sII}$ is ± 0.3 $kT_m$. An increase in the graphite content from 0 to 0.5% leads to an increase in $mQ_{sII}$ from 1.3 to 2.3 $kT_m$. The SPS activation energy for WC + C ceramics at Stage II $Q_{sII}$ varies from 4.9 to 6.9 $kT_m$ (124-175 kJ/mol) at $m = 1/3$ [34]. The effective activation energy for sintering WC + SiC ceramics is $mQ_{sII}$ ~2.0-2.2 $kT_m$. The SPS activation energy for WC + SiC ceramics at Stage II is $Q_{sII}$ = 6.0-6.6 $kT_m$ (152-167 kJ/mol) at $m = 1/3$ [37].

Thus, the SPS activation energy for sintering WC + C and WC + SiC + 0.3% C ceramics is lower than those of the activation energy of grain boundary diffusion of the carbon in α-WC (240 kJ/mol ~ 9.1 $kT_m$ [40]).

Figures 17 show $\ln(T\partial\varepsilon/\partial T) - T_m/T$ dependences for the Group II WC powder compositions with varying graphite content (Fig. 17a) and Group III WC + SiC + 0.3% C powder compositions (Fig. 17b) under the CPS dilatometry studies. The $\ln(T\partial\varepsilon/\partial T) - T_m/T$ dependences have a two-stage character. The effective CPS activation energy is $mQ_{sII}$ ~ 3.6 $kT_m$ for the WC binderless samples. Similar effective sintering activation energy values ($mQ_{sII}$ ~ 3.6-4 $kT_m$) are observed for the WC samples with 0.3 and 0.5% wt. graphite. The CPS activation energy values are 10.8-12 $kT_m$ (272-303 kJ/mol) at $m = 1/3$ and close to the table value of the activation energy of grain boundary diffusion of carbon in α-WC ceramics.

It is assumed that low values of the SPS activation energy in the stage of intensive shrinkage of the powder composition (stage II) mainly caused by the effect of the applied stress on the intensity of rearrangement and slippage of WC nanoparticles.

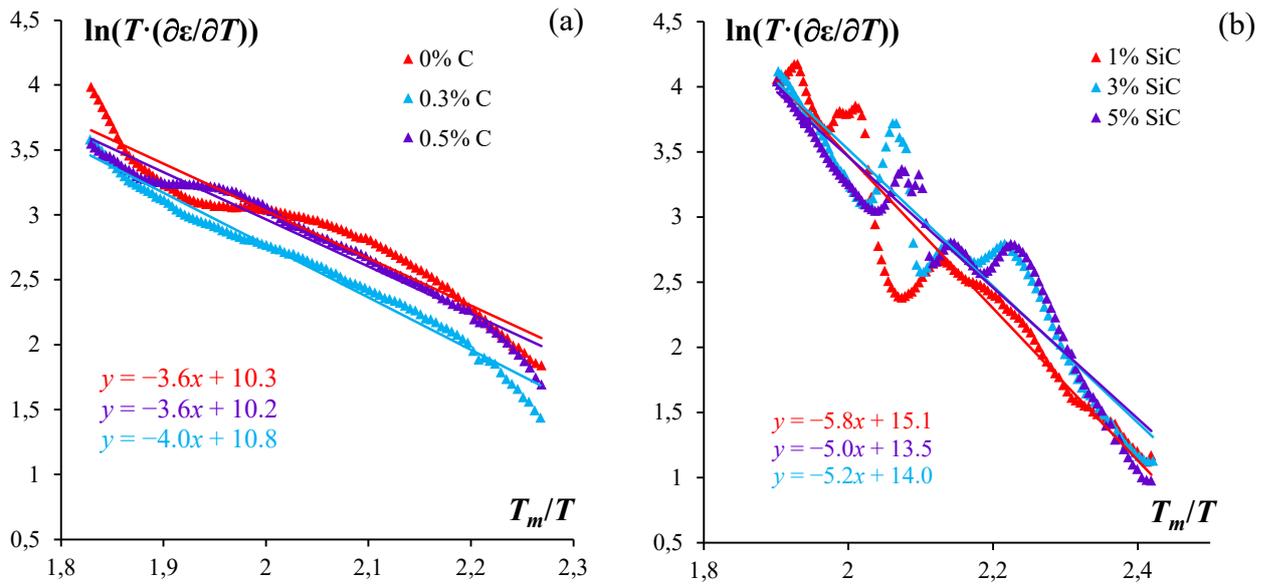

**Figure 17.** Temperature-shrinkage curves for WC + C (a), WC + SiC (b) powders in $\ln(T\partial\varepsilon/\partial T)$ – $T_m/T$ coordinates. CPS activation energy calculation for Stage II. Dilatometric data analysis

The effective CPS activation energy values for WC + 0.3% C + SiC ceramics are $mQ_{sII} \sim 5.0$-$5.8\ kT_m$. The CPS activation energy for WC + 0.3% C + SiC ceramics is $Q_{sII} = 15.0$-$17.4\ kT_m$ (379-439 kJ/mol) at $m = 1/3$. These values are close to those values of the activation energy of the $^{14}$C volume diffusion in the α-WC. At the same time, the dilatometric curves $\ln(T\partial\varepsilon/\partial T)$ – $T_m/T$ for WC + SiC + C ceramics (Fig. 17b) are more complex. Therefore, it is difficult to determine the CPS activation energy.

The slope of $\ln(T\partial\varepsilon/\partial T)$ – $T_m/T$ dependence becomes negative at high heating temperatures (Figs. 16, 17). Consequently, other approaches are required to estimate the SPS activation energy for WC-based ceramics at Stage III ($Q_{s3}$).

The activation energy at Stage III can be estimated using the diffusion model of dissolving pores located along grain boundaries in UFG materials [26]. The applicability of this procedure for WC was previously demonstrated in [14,25]. The sintering activation energy $Q_{sIII}$ is determined by the slope of $\rho(T)/\rho_{th}$ dependence in double logarithmic coordinates $\ln(\ln(\alpha\cdot\rho/\rho_{th}/(1-\rho/\rho_{th})))$ – $T_m/T$, where $\alpha = 0.33$ – compaction coefficient of α-WC nanopowders (Figs. 18) [26]. The average accuracy of determining the activation energy $Q_{sIII}$ is $\pm 1\ kT_m$.

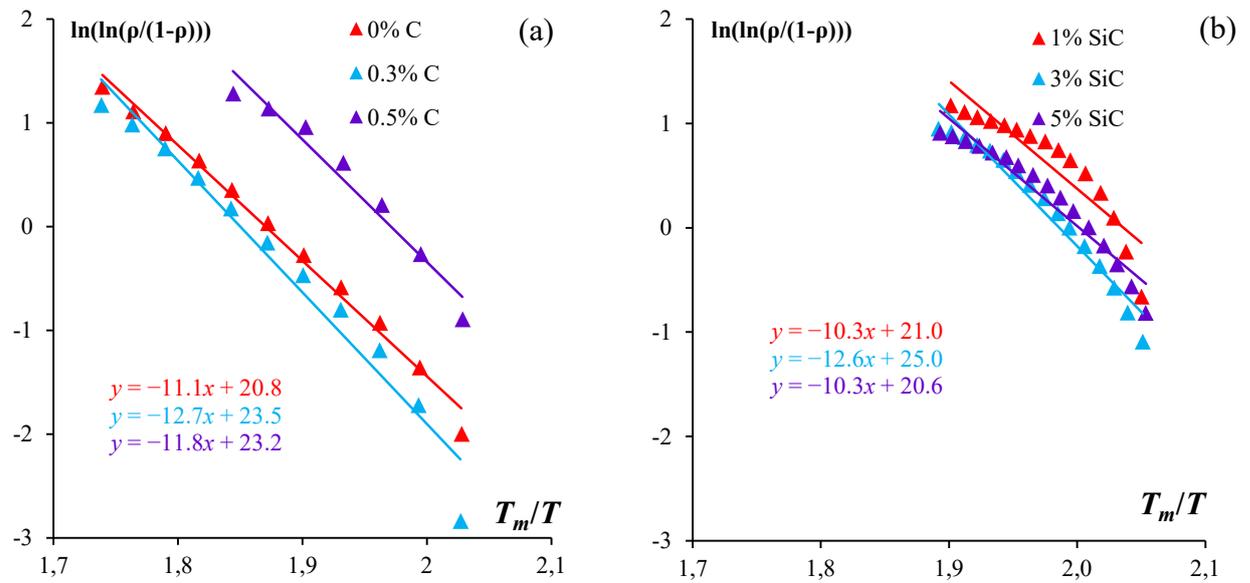

**Figure 18.** Compaction dependences on heating temperature in $\ln(\ln(\alpha\rho/\rho_{th}/(1-\rho/\rho_{th})) - T_m/T$ coordinates for the WC ceramics with varying graphite content (a) and for the WC + SiC + 0.3% C ceramics with various SiC content (b). SPS activation energy calculation at Stage III

$\ln(\ln(\alpha\rho/\rho_{th}/(1-\rho/\rho_{th})) - T_m/T$ dependence for WC at Stage III can be estimated by a straight line (Fig. 18). The addition of β-SiC particles into the WC nanopowder leads to a shift in the above dependences towards higher temperatures. At the same time, the character of dependences (slope and monotonicity) does not change with an increase in the SiC content. The SPS activation energy for WC + SiC samples is $Q_{sIII}$ = 10-12 $kT_m$ (253-304 kJ/mol) (Table 2). These values turn out to be close to those of the $^{14}$C volume diffusion in α-WC (~ 355 kJ/mol [40]). One of the reasons for increasing the SPS activation energy and changing the sintering mechanism may be the intensive grain growth in WC with an increased graphite content (Fig. 13).

As follows from Fig. 18, a two-stage character of $\ln(\ln(\alpha\cdot\rho/\rho_{th}/(1-\rho/\rho_{th})) - T_m/T$ dependence is observed for WC + SiC ceramics. The SPS activation energy for WC + SiC ceramics at Stage III-1 in the low heating temperatures range is $Q_{s3(1)}$ ~ 10-13 $kT_m$, which is close to the carbon volume diffusion activation energy in α-WC. The SPS activation energy decreases to $Q_{s3(2)}$ ~ 5-6 $kT_m$ at higher temperatures at the final sintering stage (Stage III-2). It is assumed that the SPS activation energy reduces due to a decrease in the $W_2C$ phase. The presence of $W_2C$ can cause slow down the WC ceramics sintering process.

## 5. Conclusions

1. Preliminary pressing (350 MPa) of the tungsten carbide (WC) nanopowders, reduction of heating rates from 100 to 10 °C/min as well as degassing at 950 °C are shown to lead neither to an increase in density nor to changes in the microstructure of the WC ceramic samples obtained by SPS. Fine-grained binderless WC ceramics with relative density of ~99% were obtained by heating nanopowders at 10 °C/min to 1520 °C (with exposure at this temperature for 3 min). The ceramic is characterized by high mechanical characteristics: hardness $H_V$ = 28.7 GPa and fracture toughness $K_{IC}$ = 5.0 MPa·m$^{1/2}$.

2. The shrinkage rate at the intensive shrinkage stage of SPS increases with an increase in the graphite concentration in the WC nanopowder compositions. At the same time, the WC ceramics with an increased graphite content (0.5% wt.) demonstrate abnormal grain growth. This leads to a noticeable decrease in the mechanical characteristics of the ceramics. It is optimal to add 0.3% of graphite to the WC nanopowders. This reduces the intensity of the formation of the undesirable $W_2C$ particles and ensures high mechanical properties: hardness up to 28.8 GPa and fracture toughness (Indentation Fracture resistance) up to 3.6 MPa·m$^{1/2}$.

3. Ceramics of the WC + SiC + 0.3% C composition with a high relative density (95.4-98.1%) were obtained by SPS. The ceramics have a homogeneous ultrafine-grained (UFG) microstructure with a grain size of 0.1-0.2 μm and improved mechanical properties. UFG WC + 1% SiC + 0.3% C ceramic samples have the following characteristics: hardness $HV$ = 19.1 GPa, $K_{IC}$ = 5.2 MPa·m$^{1/2}$, relative density 97.4%. The addition of SiC particles into the WC nanopowder reduces the maximum shrinkage temperature during SPS by ~150 °C and allows to reduce the volume fraction of abnormally large grains. The compaction kinetics of WC + SiC + 0.3% C powders has a three-stage character. The powder sintering kinetics at Stage II (medium temperatures) and Stage III (high temperatures) is determined by grain boundary and volume carbon diffusion in α-WC, respectively.

**Author Contributions:**

**E.A. Lantsev & M.S. Boldin** – Investigation (SPS); **P.V. Andreev** – Investigation (XRD), Data curation, Writing - review and editing, Project administration, Funding acquisition; **A.V. Nokhrin & V.N. Chuvil'deev** – Formal analysis, Methodology, Writing - original draft preparation, Writing - review and editing, Supervision; **Yu.V. Blagoveshchenskiy & N.V. Isaeva** – Investigation (DC arc plasma chemical); **K.E. Smetanina** – Investigation (XRD); **A.A. Murashov** – Investigation (SEM); **G.V. Shcherbak** – Investigation (Hardness, Fracture toughness); **K.E. Smetanina** – Investigation (XRD analysis); **N.Yu. Tabachkova** – Investigation (TEM).

**Funding:** This research was funded by the Russian Science Foundation, grant number 22-73-10175. TEM microstructure studies were carried out using the equipment provided by the Center Collective Use "Materials Science and Metallurgy" of the National University of Science and Technology "MISIS" with the financial support from the Ministry of Science and Higher Education of the Russian Federation (grant number 075-15-2021-696).

**Conflicts of Interest:** The authors declare that they have no known competing financial interests or personal relationships that could have appeared to influence the work reported in this paper.